\documentclass[manuscript,12pt]{aastex}
\def\simless{\mathbin{\lower 3pt\hbox
     {$\rlap{\raise 5pt\hbox{$\char'074$}}\mathchar"7218$}}}   
\def\simmore{\mathbin{\lower 3pt\hbox
     {$\rlap{\raise 5pt\hbox{$\char'076$}}\mathchar"7218$}}}   
\def\4u{4U 1728--34}

\received{}
\accepted{}
\journalid{337}{15 January 1989}
\slugcomment{Submitted to ApJ, March 21, 2007}

\lefthead{et al.}
\righthead{\4u}

\begin{document}

\title{Fourier resolved spectroscopy of \4u: New Insights
 into Spectral and Temporal Properties of Low-Mass X-ray Binaries}

\author{ }
\author{C.R. Shrader$^{1,4}$, P. Reig$^{2,3}$ and D. Kazanas$^1$}

\vskip 30pt
 \affil{$^1$ Astrophysics Science Division, NASA Goddard
Space Flight Center, Greenbelt, MD 20771}
\affil{$^2$IESL, Foundation for Research and Technology, 711 10
Heraklion, Crete, Greece}
\affil{$^3$Physics Department, University of Crete, PO Box 2208, 710
03 Heraklion, Crete, Greece}

\affil{$^4$ University Space Reseach Association, 10211 Wincopin Circle, 
Suite 620, Columbia, MD 21044}

\email{Chris.R.Shrader@gsfc.nasa.gov, pau@physics.uoc.gr,
Demos.Kazanas-1@nasa.gov}

\begin{abstract}

\baselineskip=15pt

Using archival RXTE data we derive the 2-16 keV Fourier-resolved spectra of
the Atoll source  \4u in a sequence of its timing states as its low QPO
frequency spans the range between 6 and 94 Hz. The increase in the QPO
frequency accompanies a spectral transition of the source from its island
to its banana states.  The banana-states'  Fourier-resolved spectra are 
well fitted by a single blackbody component with $kT \sim 2-3$ keV depending on 
the source position in the color -- color diagram and the Fourier frequency, 
thus indicating that this spectral component is responsible for the source 
variability on these timescales. This result is in
approximate agreement with similar behavior exhibited by the Z sources,
suggesting that, as in that case, the bounday layer --  the likely source
of the thermal component -- is supported by radiation pressure.
Furthermore, it is found that the iron line at $\sim$6.6 keV, clearly
present in the averaged spectra, not apparent within the limitations of our 
measurements in the
frequency-resolved spectra irrespective of the frequency range. This would
indicate that this spectral component exhibits 
little variability on time scales
comprising the interval $10^{-2}-10^2$ seconds. 
In the island state the single blackbody model proved inadequate, 
particularly notable in our lowest frequency band ($0.008-0.8$ Hz). 
An absorbed powerlaw or an additive
blackbody plus hard powerlaw model was required 
to obtain a satisfactory fit. Statisics do not allow
unambiguous discrimination between these possible scenarios.

\end{abstract}


\baselineskip=15pt

\section{Introduction}

Rapid X-ray variability and spectral distributions are powerful
probes of the physics of accretion flows onto compact objects, i.e.
onto neutron stars  and black holes. More than two decades of
persistent efforts to probe and  comprehend the physics involved in
these flows led to the accumulation of a wealth of data of both
their spectra and time variations.  Significant progress in both
these directions has been made over the last decade, in particular,
with the wealth of data provided by the Rossi X-Ray Timing
Explorer (RXTE). Its large collecting area and high time resolution
have facilitated detailed studies on the rapid variability and the
spectra  of a large sample of sources. However, the accompanying
improvement in the  understanding of the dynamics and radiation
emission physics associated with these accretion powered sources has
progressed considerably slower than the accumulation of the data and {
their ensuing classification within  semi-coherent phenomenological models.}

Concerning the differentiation between the spectral and timing
characteristics of  accreting neutron stars and black holes, the 
former appeared to present higher complexity in
both their spectral and  timing properties than the latter. This
additional complexity is generally attributed to the presence of the
neutron star surface which can complicate the dynamics of accretion
and as a result the ensuing spectral and timing properties. In the
temporal domain, the power density spectra (PDS) of neutron star
low-mass X-ray binaries (LMXBs) exhibit a variety of features
ranging from narrow  quasi-periodic oscillations (QPOs) to broad
noise components (for a review, see Wijnands  (2001); also van der
Klis (2006) and references therein), only some of which appear to
have corresponding features in the black hole case.
In the spectral domain,  while Comptonization is presumably the predominant
process for producing the  observed X-ray emission above $\sim 1$ keV, the
neutron-star LMXB spectra are quite complex, more so than those of
accreting black hole candidates. This  is likely due to the fact that a
number of components contribute to the emission in  this energy range and
that their individual spectra are hard to disentangle.

A simplified way to classify the (apparently complex) spectra of
these sources   has been that of the color-color (or color -
intensity) diagrams (Hasinger \& van der Klis 1989); these exhibit,
instead of a detailed spectral decomposition, the ratios of fluxes
in adjacent bands (colors) within the detector's range (color -
color diagrams) or one color as a function of the total flux of the
source (color - intensity diagrams). The large body of accumulated
data then made it clear that in response to changes in luminosity,
these sources cover a certain trajectory in this color - color (or
color - intensity) space whose shape has since been used to classify
them: The most luminous LMXBs, $L_x \sim 10^{38}$ erg s$^{-1}$,
cover a Z-shape track in color - color space, hence the name Z-sources. 
The less luminous sources, $L_x \sim 10^{37}$ erg s$^{-1}$, tend to
cover a circular-like path in same space, leading to their
designation as atoll sources. At the same time, it was found that
there exists a correspondence between their spectral and timing
properties, with the frequencies of timing features, such as QPOs or 
breaks, generally increasing in response to increases in the 
X-ray flux and presumably the accretion rate.

The physics governing the spectral and timing properties of Z or
atoll sources and their differences remains obscure, although as
early as 1984, Mitsuda et al. (1984) noticed the  possible
decomposition of their spectra into hard and soft components with
the hard one being the more variable of the two. A great advance in
the understanding of these spectra was made recently with  a
refinement of the arguments put forward by Mistuda et al. (1984).
This involves the implementation of a novel technique that combines the
spectral and timing properties of these sources in an effort to
produce a coherent picture of the dynamics of these sources.

This technique is known as Fourier Resolved Spectroscopy (FRS) (e.g.
Revnivtsev et al. 1999; Zycki 2003; Papadakis et al. 2005, 2006). In short,
instead of producing the power spectra as a function of photon
energy, as it is customary, this technique accumulates the
variability amplitudes over several well defined frequency bands for
each energy bin to produce the energy spectra of a source for the
specific frequency bands considered. This particular method of
presenting the data facilitates the identification of the variable
spectral components as well as the time scales over which the
variability occurs; as such, it allows certain immediate insights
into the dynamics responsible for the emission of the specific
spectral components. It was pointed out that this method is
particularly suited for  the study of spectral features that result
from X-ray reprocessing where the light crossing time provides a
natural frequency filter; in this case FRS can provide
straightforward insights about the geometry of the reprocessing
medium.  So far, this technique has been applied successfully to
galactic black hole candidates (Revnivtsev et al. 1999; Gilfanov et
al. 2001; Sobolewska \& Zycki 2006; Reig et al. 2006), to
neutron-star LMXBs (Gilfanov \& Revnivtsev et al 2005) and also to
AGN (Papadakis  et al. 2005, 2006) yielding insights into
variability of different spectral components and infer their spatial
arrangements.

Recently, the same technique has been applied to the spectro -
temporal  data of accreting neutron stars and more specifically
to those of Z-sources  (Gilfanov et al. 2003; Revnivtsev \& Gilfanov
2006). This analysis did confirm the earlier tentative decomposition
of the Z-source spectra into a variable hard component and a soft
less variable one. The normalization of these components varies with
the  accretion rate as do their spectra to produce the well known
color-color diagram of the Z-sources. In particular, the softer of
the two components  was identified with the accretion disk and the
harder one with the boundary layer of these sources.
The temperature of the former was found
to change slowly in response changes in luminosity, as
expected. The temperature of the boundary layer component remained
constant, independent of luminosity and of Fourier frequency. This
supports the conjecture that the boundary layer is dominated by
radiation pressure. It was also found (Revnivtsev \& Gilfanov 2006)
that the disk component contributes to the spectra only at the
lowest Fourier frequencies. This is consistent with the findings of
Reig et al. (2006) pertaining to the lack of variability of the disk
component around a transient black hole LMXB 4U 1543-47. This
separation and the identification of these two components in the
spectra  was possible only because of the combined spectral - timing
analysis.

In the present paper we extend the application of the method of Fourier
Resolved  Spectroscopy to the spectra of a source  of the atoll class (i.e.
neutron stars accreting at rates $\sim 10$ times lower than  those of
Z-sources). The trajectories of these sources in the  color-color diagram
have a different shape  from those of the Z-sources as discussed above. They are
known to produce a sequence of states of increasing luminosity known
as island, lower banana and upper  banana states. Variations in the spectral
properties between these states are accompanied by  corresponding changes
in their timing properties  (van der Klis 2006). The source under study is
4U 1728--34, a neutron star LMXB system  that is considered to be one of
the proto-typical examples of an atoll source. It exhibits variability on a
variety timescales, from months down to milliseconds (e.g.  Di Salvo et al
2001) as well as exhibiting Type-I bursting behavior.

In \S 2 we outline the details of our data selection and reduction
procedures, while in \S 3 we describe the FRS methodology. In  \S 4 we
present our results for \4u  offering our interpretation of the variability
of individual spectral components. In \S 5 we present our general
conclusions,  including insights gained into accreion geometry and overall
physical system configuratons associated with the separate states.

\section{Observations and Data Reduction}

Data obtained with RXTE can be collected and telemetered to the
ground in many different ways depending on the intensity of a source
and the spectral and timing resolution desired. The specific
observational modes are selected by the observer and may change
during the overall observation. This project requires a temporal
sampling of at least twice the highest frequency band to be probed,
and enough spectral resolution to separate out the different
continuum components as well as to crudely study possible iron line
signatures. Since this is an archival study, we had to examine the
various datasets for the epochs of interest in order to identify
suitable data acquistion modes. Typically, we found that the
event-by-event data sampled at 125 microseconds over 64 onboard
spectral channels covering the full bandpass of the PCA, suited our
purposes. Practically, in order to ensure homogeneity in the
reduction process and for signal-to-noise considerations the 
energy resolution was restricted to be between
16--24 channels covering the energy band 2-16 keV.

Light curves were extracted for each onboard channel range using the
current RXTE
software\footnote{http://heasarc.gsfc.nasa.gov/docs/software/lheasoft/},
binned at a resolution of 0.0078125 s. We then divided the data into
256-s segments and, following the prescription of Reig et al (2006), we
obtained the Fourier resolved spectra of the source in the following broad
frequency bands:  0.008-0.8 Hz, 0.8-8 Hz and 8-64 Hz.
In total, we examined a total of approximately 54.3 ks of data from the RXTE
archives at 7 epochs (see Table \ref{log}).

\subsection{Spectral Index -- QPO Frequency Correlation}

It is well established that in accreting neutron star systems, the
X-ray luminosity and their spectral and/or  timing quantities (X-ray
colors, QPO frequencies) tend to correlate with the spectral state
of a an individual source, but that these correlations are not identical 
for the ensemble of sources (see e.g. van der Klis 2006).
Furthermore, for a given source, the correlations are much more
pronounced over short (hours to days) than on longer time scales
(van der Klis 2000 and references therein). Therefore, the use of
hardness ratios and source intensity as tracers of the source
spectral state can become blurred. In addition, the problem of gain
changes associated with the aging of the PCU detectors aboard $RXTE$
makes even the determination and comparison of X-ray colors at
different epochs difficult.

To avoid these difficulties we have opted to use one of the QPO 
frequencies as a proxy for the state
of the source. As has been previously pointed out, e.g. van der Klis
(2006), the frequencies of all PDS features, most notably QPOs,
correlate with the source intensity and also with its spectral
state. Furthermore, there is a noted
correspondence in the sequence of the spectral and timing states
with source intensity between accreting neutron stars and black hole
systems.

Motivated by the correlation between the (low-frequency) QPO centroid
frequency and photon index in several black-hole XRBs (Vignarca et al.
2003),
we have opted to use the corresponding QPO frequency of \4u as a
tracer of the configuration changes in the accretion flow. Although
as noted above, the neutron star spectra are more complicated than
those of the black holes, we find that the correlation between the
spectral and timing properties of \4u and similar systems (van der
Klis 2006) is sufficiently robust to allow the use of the QPO
frequency as an indicator of the spectral state. In fact, Titarchuk
\& Shaposhnikov (2005) found a correlation between the QPO frequency
and their spectral index parameter $\Gamma$ in their analysis of \4u.
However, given that its
spectrum is more complicated than those of galactic black holes one
should not consider this correlation to be universal.

\section{Conventional Spectral and Timing Analysis}

Fig 1 shows typical energy spectra of the source corresponding to the 
observational periods selected. The spectra were extracted from PCA 
{\em Standard 2} mode data, and in this case no Fourier frequency 
decomposition has been applied.
The response matrix and background models were created using the standard
HEADAS software, version 5.3. The number of detectors (PCU) that were
switched on for each observation varies, and, in order to be able to
compare the spectra, we divided them by the respective number of PCUs.

The spectral analysis was performed using the XSPEC analysis package
version 11.3.1. We have added systematic errors of 1\% to all channels and
have restricted our analysis to the 2-16 keV band only (to match the energy
band used in the case of the Fourier-resolved spectra). The results are
listed in  Table \ref{avsptab0}. The errors quoted for
the best fit values correspond to the 90\% confidence limit for one
interesting parameter.


We analyzed data at seven different epochs as indicated in Table 1.
The 21 and 26
September 1997, 25 January 1999 and 9 February 2001,
observations were selected because they span a wide range in both
the QPO and break frequencies, $\nu_1$ and  $\nu_{\rm b}$, respectively,
as well as the spectral parameter $\Gamma$ of \citep{tita05}. As 
mentioned above we inferred the spectral state of \4u\
for a given observation based on the pair ($\Gamma$, $\nu_{\rm b}$).
In three cases, 7 June 2001, 18 February 1996, and 19 August, 1999
the source was known to be in a particlular region of the 
color-color diagram based on results in the published literature.

As shown in Table 2 the average spectra comprise a
number of continuum and atomic transition components: (a) A low temperature
disk blackbody of innermost temperature  $T_{\rm in} \simeq 1$ keV. (b)
A blackbody component of $T_{\rm bb} \sim 2$ keV attributed to the
boundary layer. (c) A power law with index $\Gamma \simeq 2$, presumably 
the result of
the Comptonization of the disk or boundary layer photons by the accretion
disk corona. (d) An Fe line feature at $E_{\rm c} \sim 6.6$ keV and, (e) An
Fe line edge at $E_{\rm edge} \sim 9.5$ keV. In figs \ref{is}a,b. we
present the overall spectrum as the decomposition of these three components
for two epochs when the source was in its island (09 Feb. 2001) and upper
banana (21 Sep. 1997) states. The disk blackbody component is not required
in the spectrum of the island state (IS) of Feb 09, 2001 but it is
necessary to fit the lower banana (LB) and upper banana (UB) branch spectra.  
The blackbody temperature $T_{\rm bb}$  appears to increase as the source moves
from the IS to the UB while its normalization traces the overall increase
in the source counts along the above sequence. The progression from the IS
to the LB and UB states is also accompanied by an increase of the power law
normalization and its index $\Gamma$.  The latter is usually interpreted as
a steepening of the Comptonization spectrum due to the cooling of the
Comptonizing electrons by the increased soft (blackbody)  photon flux.

 We have also found that the source variability properties change
markedly with the spectral state. First, as shown in Table 1 the
break and QPO frequencies increase monotonically as the source flux
and power law index $\Gamma$ increase. In Fig 3 we present
the PSD of the source in three different spectral states. The island
state exhibits the highest RMS variability at all frequency bands,
its $\nu P_{\nu}$ spectrum exhibiting a maximum at the limit of the
searched frequencies ($\sim 500$ Hz). In the lower banana state the
$\nu P_{\nu}$ spectrum rises from the low frequencies to a plateau
(i.e. $\nu P_{\nu}\propto \nu^0$ at $\nu \sim 20$ Hz) and the RMS
variability is lower at all frequencies than that of the island
state. Finally, in the upper banana state the shape remains
qualitatively similar to that of the LB but with greatly reduced
amplitude and the additional appearance of excess power at
frequencies lower than 0.3 Hz. We believe that this sequence in the
timing properties is suggestive of the progressive influence of the
effects of radiation pressure in the accretion process (negligible
in the IS but progressively important towards the UB state). It 
is reasonable to consider the possibility that 
the radiation feedback on the flow could 
damp the short scale variability by the diffusion of photons through
the accreting plasma;  inundating the system with soft photons 
that could conceivably also reduce the overall variability 
amplitude \citep{kazhua}. 

\section{Fourier Resolved Spectral Analysis}

In this section we present the results of our Fourier resovled spectral 
analysis for each state and compare the resulting best-fit
parameters to those of the previous section (i.e, those obtained from the
average energy spectra).  As before, the { XSPEC} package was used for
the model fitting. In all cases we added an absorption component which
we fixed at $3\times10^{22}$ cm$^{-2}$ (e.g. Narita, Grindlay \& Barret, 1999).
A uniform systematic error of 1\%
was added quadratically to the statistical error of all Fourier spectra in
each energy channel. Errors quoted for the best-fit values correspond to
the 90\% confidence limit for one interesting parameter.

The virtue of the frequency-resolved spectra is that they receive
significant contribution only from the spectral components that  are
variable on the time scales sampled; the FRS do not
represent photon rates but rather variability amplitudes at a given
energy. Therefore, through Fourier-resolved spectroscopy we can
investigate whether different spectral components in the overall
spectrum of the source (e.g. blackbody, Comptonization, iron line)
are variable and at which frequency.  In general, the interpretation
of the FRS is not unique (see Papadakis et al. 2005,
Reig et al. 2006 for some insights). For spectral features
that result from the reprocessing of higher energy ($E\ge 7$ keV)
continuum, such as the iron Fe K$\alpha$ line and the Compton
reflection "hump", one plausible interpretation of
the Fourier-resolved analysis is that the light crossing delay $R/c$ 
(where $R$
is the size of the reprocessing region) filters out variability of
these features on frequencies higher than $\sim c/R$. We would note that
while this interpretation may be the most straightforward one, 
other interpretations, such as scenarios 
involving screening geometries are also possible e.g. (\cite{revn99}).

In the following subsections we describe the results of our 
analysis for each of the spectral
states of the source. In addition to the four sets of data mentioned
above, we analyzed data for three additional epochs for which the spectral
state of the source was previously documented. Our objective is to
compare the results obtained above with the traditional designations
of island and banana states in order to characterize the spectral
and timing properties of each state. We searched in the literature
for observations for which \4u\ was clearly identified as being in
one of the characteristic atoll states. \4u was in the island state
on June 7, 2001 \citep{migl03}, upper banana state on August 19,
1999 \citep{pira00} and lower banana state on February 18, 1996
\citep{salv01}.

\subsection{The Island State}

The source was in the island state on Feb. 09, 2001, and, as noted by
Migliari et al. (2003) \4u\ also appeared to be in the island state during
the June 7, 2001 observations. We obtained the 64-channel, 1-ms data sets 
from those observations and formed the FRS as described in \S 3. 
We found that a single-temperature blackbody model led to fits which were
unacceptable, particularly at our low-frequency band. In particular, 
residuals at the low- and high-energy channels were evident.
Improved fits were provided with an additive combination of the blackbody 
plus a powerlaw. The powerlaw required is rather 
flat $\Gamma \simeq 0.5$, but is is poorly constrained, particulary
in the  Feb. 09, 2001 observation. The reason is that in those model
fits the blackbody is the predominant component except in the 
(several) highest energy energy bins. 

The blackbody component in these cases is characterized at each frequency 
band by a temperature of about $kT=1.0-1.7$ keV without an obvious trend. 
The spectral decomposition parameters for the island-state FRS
are given in Tables \ref{ISplfits} and \ref{ISbbplfits}. As summarized 
there and shown in fig. \ref{island}, the Feb. 9, 2001 FRS, for example,
can be fitted well by a 
blackbody plus powerlaw spectrum with temperature $k T \sim 1.5-1.7$ keV, 
similar in magnitude to the temperature $T_{\rm bb}$ of the conventional
spectra attributed to emission by the boundary layer. { The Power Law 
(energy) spectral indices} are quite hard ($\simeq 0.6$) although they are 
marginally constrained. The same spectra can actually be fit by a 
simple power law absorbed at low energies, without the presence of a 
black body component. A summary of such fits is given in Table \ref{ISplfits},
for an absorbing column of $N_H = 3 \times 10^{22}$ cm$^{-2}$. Also
see \ref{island}. The fits
exhibit no particular trend with Fourier frequency. 

We also tested several other simple models. 
A powerlaw model with low-energy cutoff was unable to 
constrain the inflection energy value and broken powerlaw fits
led to unacceptable results. 

There seems to be a general trend in the island state FRS 
towards hard or even positively sloped spectral-energy 
distributions, i.e. a trend
variablity  to increase with energy. This is particularly notable
for example in the low-frequency spectra for both island state epochs
(Tables \ref{ISplfits} and \ref{ISbbplfits}), and it is consistent
 with the RMS variability versus energy analysis of \cite{revn00}.
This is reminiscent of Blazar variability.

While none of the FRS results require an iron line - upper limits
on its equivalent width range from a few hundred to a thousand 
eV - it is clearly
seen in the conventional spectra for 3 of the 4 epochs
denoted in Table 2; specifically the EW determination indicates
a $\sim 10\sigma$ significance in several cases. The FRS
are harder than the averaged spectra, indicating again that the
hard emission may be the predominant component of the source variability.

Finally, we note that although the island-state count rate was lower
by about a factor of 10 than the banana-state cases, the rms 
variability in the island state is significantly higher. Thus, 
the statistics of the island-state FRS are comparable or 
superior to those of some banana state epochs.

\subsection{The Lower Banana state}

On Jan. 25, 1999 as well as on Feb. 18, 1996 (also see di Salvo et
al. 2001) the source intensity along with the PDS break frequency at
about 16 Hz, indicated that \4u\ was in the the lower banana state.
A single blackbody model provides acceptable fits for all three of
our frequency ranges (our fit results are summarized in Table
\ref{frstab1}). Its temperature is consistent with that of the
time averaged spectra, indicating that it is the boundary layer
component which is responsible for the source variability. The
absence of the power law and Fe line components that are present in
the time averaged spectral fits, suggests that these components are
not variable on the time scales examined. The normalization of this
component is comparable in the low and high frequencies and slightly
lower at the mid-frequency range, indicating that all frequencies
contribute roughly equally to the variations of this component.

The single blackbody fit to the FRS for various banana spectral states,
and for the three frequency bands have characteristic temperatures
that span the nominal kT$\simeq 2-3$ keV range. However, no 
unambiguous trends
with temperature versus frequency are seen. 

%
%

\subsection{The Upper Banana State}

The source was found to be in the upper banana state on Sept.
26 and Sept. 21 1996, as determined by the value of the PSD break
frequency $\nu_b$.  We also obtained data for the upper banana state
noted during August 19, 1999 by (Piraino et al. 2000). In this case,
we found that the RMS variations above $\sim$8~Hz (high-frequency)
were apparently small as we were unable to obtain useful
signal-to-noise for any of our FRS bands for these data. The overall
count rate on the other hand was high, $\sim 10$ times that for the
island state. As discussed above a single blackbody provides a good
fit to the FRS without the need for additional components at
any of the frequency bands. The amplitude of this component appears
to increase with decreasing frequency in the 1996 observations while
remained the same during the 1999 one, thus indicating that this
attribute of the spectra is not characteristic of the particular
state.

\section{Discussion}

We have investigated the spectro-temporal characteristics of 
the atoll X-ray binary \4u. To this
end we have produced the energy and power spectra of the source in
its different spectral states, i.e. island (IS), lower banana (LB) and
upper banana (UB) and also the Fourier-resolved spectra (FRS) of each 
state for three different frequency bands.

Fits of the time average spectra of the source require a combination of a
blackbody (to model the boundary layer emission), a multicolor blackbody
(to model emission by the accretion disk), a power law (to model coronal
emission), an Fe line and a low-energy galactic-absorption 
column of $N_H = 3\times 10^{22}$ cm$^{-2}$. The results 
for the conventional spectra
are given in Table \ref{avsptab0}. As the source crosses from the island to
the lower and upper banana states the source spectra change significantly
(see Fig. 1). Our spectral decomposition indicates for
this sequence the component exhibiting the largest parameter
range is the multi-color disk. While its presence is not required 
in the island state, it is necessary for acceptable fits in the other states 
with approximately
constant temperature and highly variable normalization. Unfortunately, the
RXTE spectra do not cover the energy regime $E < 2$ keV, necessary for the
more accurate determination of this component. Also variable is the index
$\Gamma$ of the power law component; the increase of $\Gamma$ with the
total flux of the source is generally attributed to the decrease in the
temperature of the Comptonizing medium by the increase of soft thermal
photon flux and the resulting Compton cooling effects. The temperature of the 
blackbody component exhibits an increase between the island and the banana
states (referring to the blackbody plus powelaw fits to the island state), 
but it appears to remain constant once the banana states are
reached. This suggests a behavior different from that of the Z-sources.
Gilfanov et al. (2003) find from their FRS analysis
that the temperature of this component remains constant, 
suggesting that
the sources' boundary layer is radiation pressure dominated. Following the
same argument, the increase of the thermal component temperature with
increasing source flux leads to the conclusion that in the atoll sources
the boundary layer is not radiation dominated  in the island state, while
it appears to become so in the banana states. Finally, the EW of the Fe
line appeared to remain constant to within statistics throughout the source's 
spectral sequence.

The variability properties of the source, as manifest by the source 
power-density spectra, change significantly with its spectral
state: The overall RMS fluctuations across all frequencies decrease with
increasing source flux; in addition the PDS shape changes as the source
moves from the island to its banana states (see Fig. 3). Finally,
in the general case, the frequency of features in the PDS such as breaks
and QPOs increase along the island to upper banana sequence. We suggest
that the decrease in variability is due to the effects of radiation
pressure on the accretion flow, which through diffusion damps the high
frequency components and the overall amplitude. 

We have elaborated further on the study of the spectro-temporal
properties of the source by producing its Fourier-resolved spectra
at three frequency bands for each of the source's spectral states,
namely $0.008 - 0.8$ Hz, $0.8 - 8$ Hz and $8 - 64$ Hz. While
interesting features in the Fourier domain are known to occur at
higher frequencies (i.e. kHz QPO) our data did not allow the
application of this technique to these frequencies with any
significance. The source state was determined either from prior
analysis gleaned from the published literature or was defined by the
spectral hardness--break frequency correlation. In the latter
cases, we mostly relied on the measurements and analysis of
\citep{tita05}, but in several instances we measured these
quantities ourselves. Several notable trends are gleaned from our
analysis and are identified below.

The results of our FRS analysis can be summarized as follows:

(i) The iron line at $\sim$6.6 keV, distinctly present in the
conventional spectra ($\simeq 10 \sigma$ for the banana state cases), 
is not apparent in any of our frequency-resolved spectra.
For example, when we fix the line energy at 6.6 keV and the width 
to 0.6 keV, we derive only upper limits on the equivalent widths
(values ranged from about 50 to 600 ev). 
This suggests that, at leaset within the limitations of our 
measurements, there may be no significant Fe line
variability on time scales { between $10^{-2} - 10^2$ s}. This
is in agreement with the lack of correlation between the Fe
K$\alpha$ equivalent width and QPO frequencies reported by
\citep{tita05}. These authors concluded that the size of the
Fe K$\alpha$ emitting region is not of the same order as that of the
X-ray continuum emission, the former possibly being orders of
magnitude larger than the latter. On the other hand, a different
view has been promoted by Miniutti et al. (2003) with respect to the
Fe line variability in AGN. In as much as the geometry and physics
of Fe line production in these very diverse objects is similar, such
an interpretation should also be considered.

(ii) No disk component is present in the FRS, indicating as in the
case of the iron line, that the accretion disk emission is significantly
less variable at the frequencies sampled by our study, i.e. 0.08-64 Hz.
This result agrees with those found in other neutron-star \citep{revn06}
and black-hole \citep{reig06} binaries.

(iii) We have
examined five data sets corresponding to the banana state
spectral/temporal configuration. Our FRS reveal that the temperature of the
thermal spectral component is consistent with being constant across
frequencies and states. Unfortunately, this statement cannot be made with
high degree of confidence for the UB states as the decrease in the variability
amplitude makes the determination of the Fourier resolved
parameters of the spectra difficult. 

%
%

(iv) Two observations corresponding to the island state reveal evidence
for a spectral hardening at the lowest frequencies examined. The island
state are generally harder than the corresponding frequency averaged
spectra. Single blackbody model fits to the island state FRS, 
were unsatisfactory, with a significant high-energy residual appearing,
particularly for our lowest frequency interval. The addition of a 
hard powerlaw component leads to improved fits to the data, although 
the slope is poorly constrained. { We have
also tried an alternative models, the most satisfactory of which was
 a single power law with a low energy 
neutral absorpion by a column of $N_H = 3 \times 10^{22}$ cm$^{-2}$. It was 
found that this presents an good fit to the data with the 
corresponding parameters as given in Table \ref{ISplfits}. In general,} 
the FRS are notably harder than the conventional spectra for the island state, 
indicating a substantial contribution of the high-energy emission to the 
source variablity, at least for this spectral state, a situation not unlike that
found in other studies as previously referenced. 

(v) The results of our FRS analysis are in general agreement with those of
Gilfanov et al. (2003) of Z-sources, in that the FRS spectra are dominated
by a single blackbody component of constant temperature when
$\dot{M}\sim 0.1-1 \; \dot M_{\rm Edd}$, that is, for Z sources and atoll
sources in the banana state. This is attributed
to emission by the boundary layer which in addition is radiation pressure
dominated hence the independence of the temperature on the source flux (the
local flux is always that of Eddington). The temperature of this component
is consistent with that of the blackbody component in the time average
spectra. 

(vi) Finally, we have extended the study of the Fourier-resolved  spectra of
neutron-star low-mass X-ray binaries to the lower luminosity end. Previous
studies had concentrated on { higher luminosity states of these objects}, 
i.e., Z-sources or atoll sources in
their soft/high spectral state \citep{gilf03,gilf05,revn06}. The X-ray
luminosity in the 2--16 keV range of \4u\ in the island state, assuming a
distance of 4.3 kpc \citep{fost86} is $\sim 3\times 10^{36}$ erg s$^{-1}$,
i.e. $\sim 10^{-2} L_{\rm Edd}$, about one order of magnitude lower than
previous studies. We found that the temperature of the blackbody component
of the conventional spectrum of the island state is significantly lower than
that of the corresponding FRS when the powerlaw plus blackbody model
is applied. The most straightforward 
interpretation of this fact is
that the conventional island-state spectrum includes
contribution from a component of lower temperature and little variability.

Future applicaton of Fourier-Resolved spectroscopy to low-mass X-ray 
binaries, incorporating both larger data sets and additional objects of the
representative subclasses offers the posiblity for furher insight 
into each of these issues.

{\noindent{\bf Acknowledgements}}

The authors wish to acknowledge Nikolai Shaposhnikov for providing
the \citep{tita05} spectral-QPO frequency correlation results in 
tabular form. We also thank the referee, Mikhail Revnivtsev for 
a thorough reading of the manuscript and a number of useful suggestions.
This work has made use of data obtained through the High 
Energy Astrophysics Science Research Center of the NASA Goddard
Space Flight Center. Part of this work was supported by the General 
Sectreteriat of Research and Technology of Greece.

%
%
\clearpage
\begin{table}
\begin{center}
\caption{Log of the observations \label{log}}
\begin{tabular}{lccccc}
\tableline \tableline
Epoch       &$\Gamma$ &$\nu_{\rm b}$ &Source    &2-16 keV   &Exposure\\
        &    &(Hz)      &state  &flux (c/s)$^*$ &time(ks) \\
\tableline
Feb 9, 2001     &1.40   &0.75       & IS+  &640    &4.40   \\
Jan 25, 1999    &2.07   &12.54      & LB+  &1824   &11.3   \\
Sep 26, 1997    &2.88   &24.2       & UB+  &1659   &12.8   \\
Sep 21, 1997    &5.54   &41.55      & UB+ &1953   &13.5   \\
\tableline
Jun 7, 2001 & 1.46  &  0.9     &IS &445    &6.91   \\
Feb 18, 1996    &2.41   &12.21      &LB &1380   &3.33   \\
Aug 19, 1999    &3.35   & 24.1      &UB &2600   &2.05   \\
\tableline \tableline
\tablenotetext{*}{For 5 PCU}
\tablenotetext{+}{Inferred from PDS}
\end{tabular}
\end{center}
\end{table}
%
\clearpage
\begin{table}
\begin{center}
\caption{Conventional Spectral Analysis Results \label{avsptab0}}
\begin{tabular}{lcccc}
\hline \hline
Observation     &Feb 9, 2001    &Jan 25, 1999       	&Sep 26, 1997   &Sep 21, 1997 \\
\hline \hline
Flux$^1$ (erg s$^{-1}$ cm$^{-2}$)   &1.4        &4.1     &3.8        	&4.6 \\
\hline
\multicolumn{5}{c}{DISKBB} \\
\hline
$T_{\rm in}$ (keV) &--      &1.7$^{+1.1}_{-0.5}$   	&1.06$\pm$0.03       	&1.4$\pm$0.1 \\
norm          &--       &5$^{+10}_{-4}$     		&110$\pm$15  		&45$^{+20}_{-10}$ \\
\hline
\multicolumn{5}{c}{BBODY} \\
\hline
$kT_{\rm bb}$ (keV) &1.27$\pm$0.05    &2.3$\pm$0.4         &2.1$\pm$0.1   	 &2.2$\pm$0.1 \\
norm ($\times 10^{-2}$)&0.34$\pm$0.08 &1.3$^{+0.5}_{-0.1}$ &2.8$\pm$0.1		&3.3$\pm$0.3 \\
\hline
\multicolumn{5}{c}{POWER LAW} \\
\hline
$\Gamma$        &1.75$\pm$0.04      &2.03$\pm$0.08    	&2.1$\pm$0.2      	&2.3$^{+0.2}_{-1.4}$ \\
norm            &0.28$\pm$0.04      &1.0$\pm$0.2      	&0.4$\pm$0.2  		&0.42$\pm$0.05 \\
\hline	
\multicolumn{5}{c}{GAUSS} \\
\hline
E$_{\rm c}$ (keV)   &6.6$\pm$0.2    &6.6$\pm$0.2      	&6.5$\pm$0.3    	&6.6$\pm$0.3 \\
$\sigma$ (keV)      &0.5$\pm$0.4    &0.6$\pm$0.2    	&0.8$\pm$0.3    	&0.7$\pm$0.3 \\
EW (eV)             &100$\pm$60    &150$\pm$15       	&180$\pm$20     	&140$\pm$15 \\
\hline
\multicolumn{5}{c}{EDGE} \\
\hline
E$_{\rm edge}$ (keV)&--     	&9.5$\pm$0.5      	&9.6$\pm$0.6    	&9.9$\pm$0.6 \\
$\tau$          &--     	&0.04$\pm$0.02     	 &0.04$\pm$0.03    	&0.04$\pm$0.03 \\
\hline \hline
\multicolumn{5}{l}{All fits include a fixed absorption component with $N_{\rm H}=3\times10^{22}$ cm$^{-2}$} \\
\multicolumn{5}{l}{$^1$: $\times 10^{-9}$ in the energy range 2--16 keV} \\
\end{tabular}
\end{center}
\end{table}
\clearpage

\begin{table}
\begin{center}
\caption{Results of the spectral fits (single power law) to the
Fourier-resolved spectra spectra of the island state. The fits include an
absorption compoonent with fixed $N_{\rm H}=3\times 10^{22}$ cm$^{-2}$. 
\label{ISplfits}}
\begin{tabular}{lcc}
\hline \hline
Observation			&09/02/01		 &07/06/2001	\\ 
\hline
\multicolumn{3}{c}{0.008--0.8 Hz} \\			
\hline
$\Gamma$			&1.7$\pm$0.1		 &1.8$\pm$0.2	 \\
norm ($\times 10^{-2}$)		&2.0$\pm$0.3		 &0.9$\pm$0.4  \\
\hline
\multicolumn{3}{c}{0.8--8 Hz} \\			
\hline
$\Gamma$			&1.7$\pm$0.1		 &1.7$\pm$0.2	 \\
norm ($\times 10^{-2}$)		&3.4$\pm$0.6		 &1.6$\pm$0.5  \\
\hline
\multicolumn{3}{c}{8--64 Hz} \\
\hline
$\Gamma$			&1.6$\pm$0.3		 &1.4$\pm$0.6	 \\
norm ($\times 10^{-2}$)		&3.0$\pm$1.5	      	 &1.0$\pm$0.8	 \\
\hline
$\chi^2$/dof			&1.5/52			&0.91/46\\
\hline \hline
\end{tabular}
\end{center}
\end{table}

\begin{table}
\begin{center}
\caption{Results of the spectral fits (blackbody plus power law) to the 
Fourier-resolved spectra spectra of the island state \label{ISbbplfits}}
\begin{tabular}{lcc}
\hline \hline
Observation			&09/02/01		 &07/06/2001 \\	 
\hline
\multicolumn{3}{c}{0.008--0.8 Hz} \\			
\hline
$kT_{\rm bb}$ (keV)		&1.7$\pm$0.1		 &1.0$\pm$0.3	 \\
norm ($\times 10^{-3}$)		&0.7$\pm$0.1		 &0.19$\pm$0.04  \\
$\Gamma$			&0.6$\pm$0.3		 &0.4$\pm$0.7	 \\
norm ($\times 10^{-3}$)		&0.55$\pm$0.06		 &0.35$\pm$0.07  \\
\hline
\multicolumn{3}{c}{0.8--8 Hz} \\			
\hline
$kT_{\rm bb}$ (keV)		&1.5$\pm$0.1		 &1.7$\pm$0.3	 \\
norm ($\times 10^{-3}$)		&1.0$\pm$0.1		 &0.5$\pm$0.1	 \\
$\Gamma$			&0.6$^f$		 &0.4$^f$	 \\
norm ($\times 10^{-3}$)		&1.4$\pm$0.1		 &0.32$\pm$0.06  \\
\hline
\multicolumn{3}{c}{8--64 Hz} \\
\hline
$kT_{\rm bb}$ (keV)		&1.7$\pm$0.6		 &1.0$^{+2.0}_{-0.5}$\\
norm ($\times 10^{-3}$)		&0.9$^{+0.9}_{-0.4}$  	 &0.9$\pm$0.3	 \\
$\Gamma$			&0.6$^f$		 &0.4$^f$	 \\
norm ($\times 10^{-3}$)		&1.6$\pm$0.4	      	 &0.8$\pm$0.2	 \\
\hline
$\chi^2$/dof			&0.90/51		&0.73/43\\
F-test prob.$^*$($\times 10^{-2}$)	&$7\times 10^{-8}$    	&$7\times 10^{-6}$\\
\hline \hline
\multicolumn{3}{l}{$^f$: fixed parameter} \\
\multicolumn{3}{l}{$^*$: probability that the addition of the power-law
component occurs by chance} \\
\end{tabular}
\end{center}
\end{table}

 \clearpage
\begin{table*}
\begin{center}
\caption{Results of the spectral fits (blackbody) to the Fourier resolved
spectra spectra  \label{frstab1}}
\begin{tabular}{lccccc}
\hline \hline
Observation         &Feb 18, 1996     &Jan 25, 1999    &Sep 26, 1997     &Sep 21, 1997	&August 19, 1999 \\
\hline
\multicolumn{6}{c}{0.008--0.8 Hz} \\
\hline
$kT_{\rm bb}$ (keV)     &2.7$\pm$0.4 &2.27$\pm$0.04   &2.27$\pm$0.04	&2.14$\pm$0.05	&2.4$\pm$0.2 \\
norm ($\times 10^{-3}$) &1.0$\pm$0.1 &4.8$\pm$0.1     &5.0$\pm$0.1	&3.7$\pm$0.1	&1.8$\pm$0.1 \\
\hline
\multicolumn{6}{c}{0.8--8 Hz} \\
\hline
$kT_{\rm bb}$ (keV)     &2.10$\pm$0.08 &2.02$\pm$0.05  &2.3$\pm$0.2	&2.1$\pm$0.4	&2.5$\pm$0.5  \\
norm ($\times 10^{-3}$) &2.6$\pm$0.1   &3.07$\pm$0.07  &1.5$\pm$0.2	&1.4$\pm$0.3	&2.2$\pm$0.4  \\
\hline
\multicolumn{6}{c}{8--64 Hz} \\
\hline
$kT_{\rm bb}$ (keV)     &2.18$\pm$0.09   &2.15$\pm$0.07  &2.2$\pm$0.4   &2.2$\pm$0.3	&2.8$\pm$0.6  \\
norm ($\times 10^{-3}$) &4.2$\pm$0.2     &4.6$\pm$0.1    &2.2$\pm$0.3   &2.5$\pm$0.3	&4.0$\pm$0.8  \\
\hline
$\chi^2$/dof            &0.95/37        &1.80/64        &1.05/53        &0.94/55	&0.95/39 \\
\end{tabular}
\end{center}
\end{table*}

\clearpage
\begin{figure}
\epsscale{0.5} \plotone{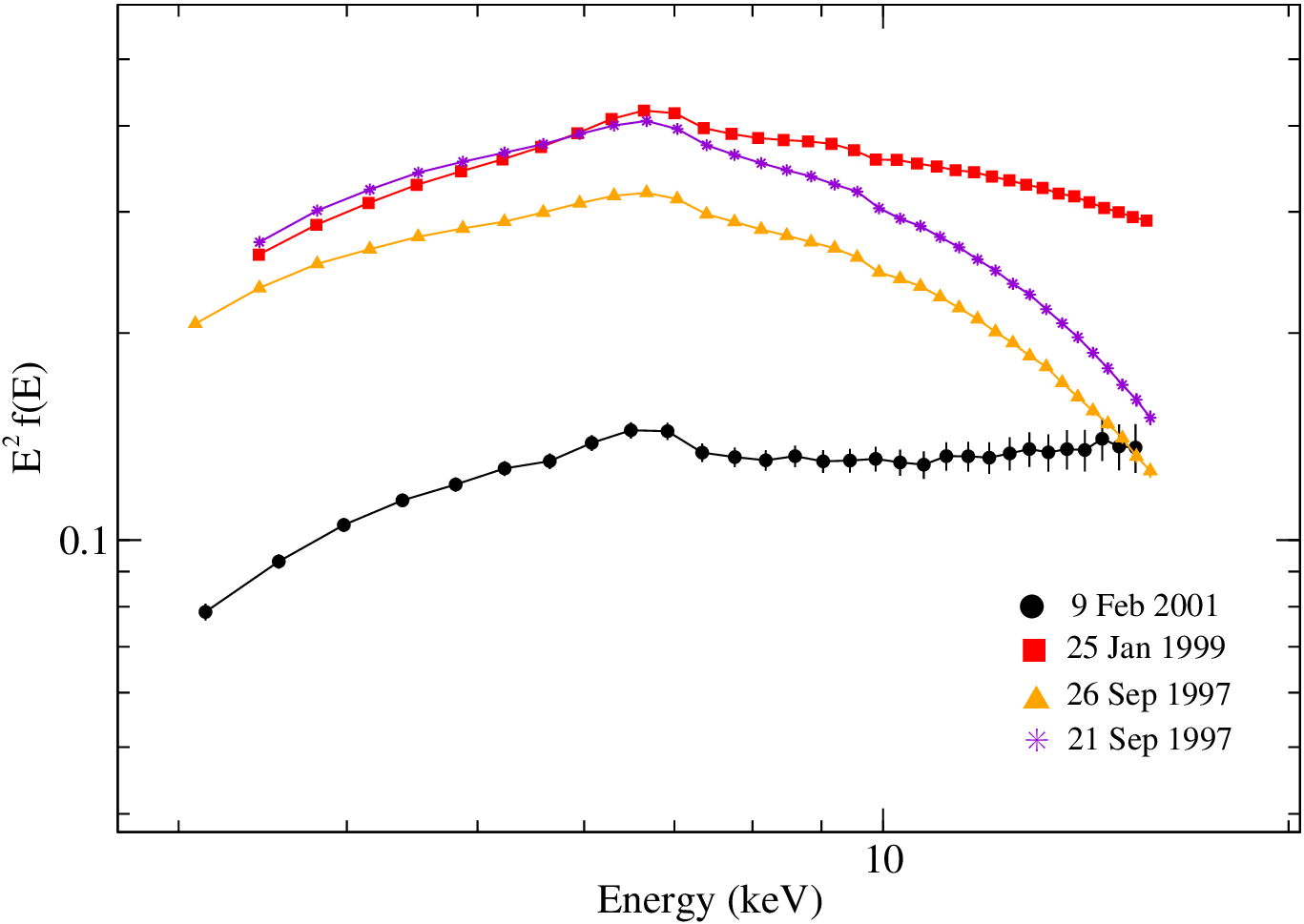} \caption{Results of our (conventional)
spectral analysis for the various epochs and corresponding spectral state
configurations. The spectra are satisfactorily fitted with an absorbed blackbody 
plus multi-color accretion disk components at low energies and a 
power-law component at higher energies. An Fe line plus edge were also
required (see Table~\ref{avsptab0}).}  \label{avsp}
\end{figure}
\clearpage

\begin{figure}
\epsscale{1.0}
\plotone{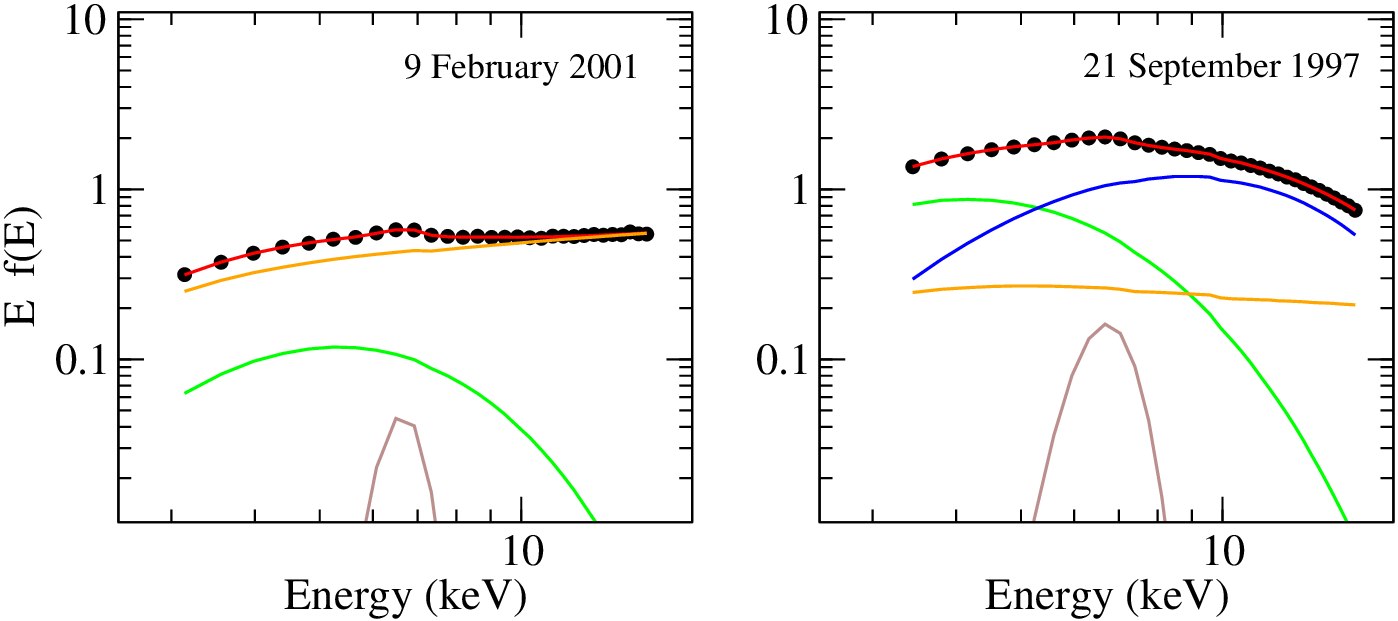} \caption{Spectral
decomposition of the average source spectra for the island (09
February 2001) and upper banana (21 September 1997) states. The model
components are described in the text and in Table \ref{avsptab0}, and 
are plotted in $\nu f_{\nu}$ space. Specifically, for Feb. 9 2001 a blackbody
component (green), powerlaw (brown) and a gaussian line (purple) 
are depicted; for Sep. 21 1997 one can also see the blackbody component (blue),
the multi-color disk (green), the powerlaw (brown) and the line (purple).
As the
source flux increases so does the blackbody temperature, the index
of the power law and the Fe line flux. } \label{is}
\end{figure}

\clearpage

\begin{figure}
\epsscale{1.} \plottwo{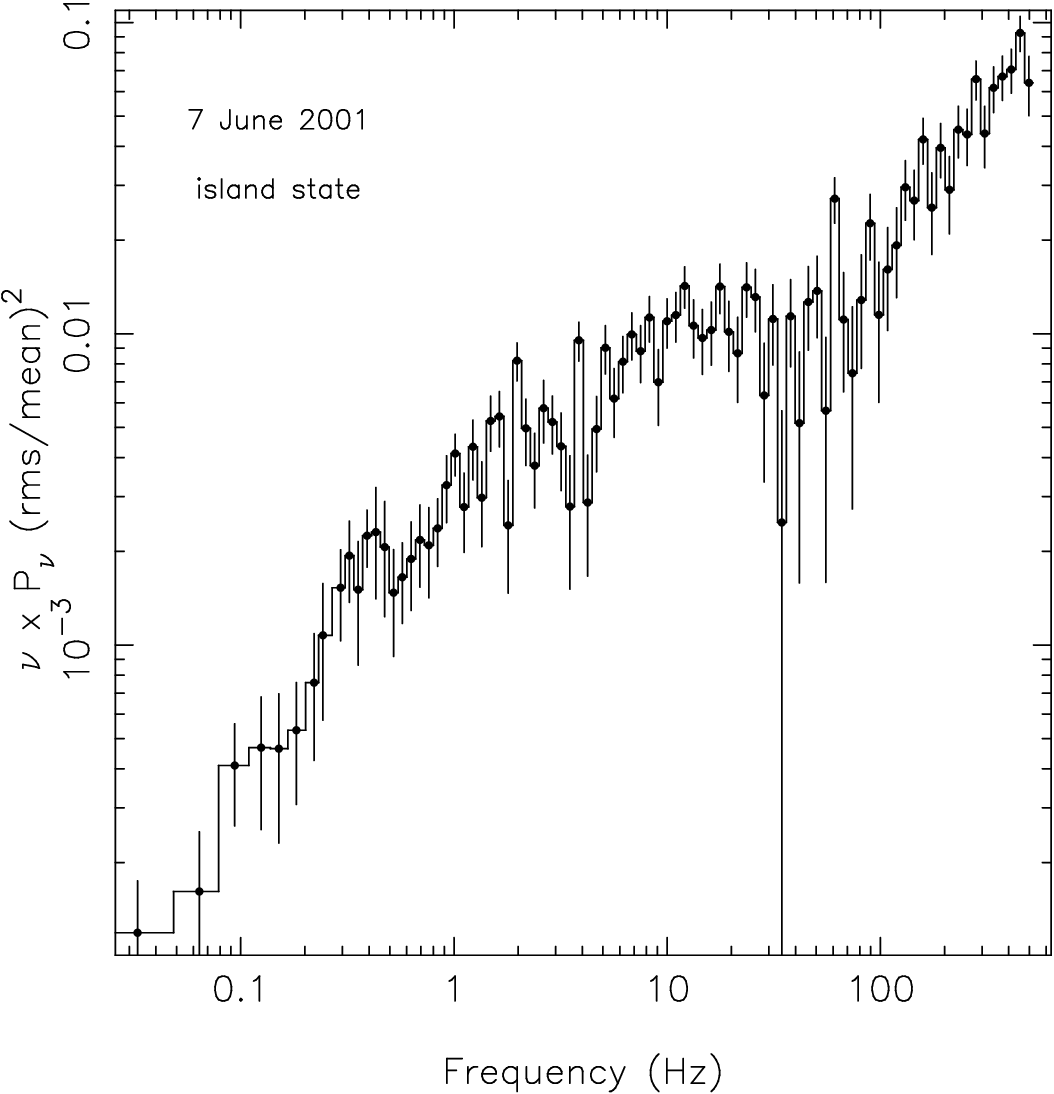}{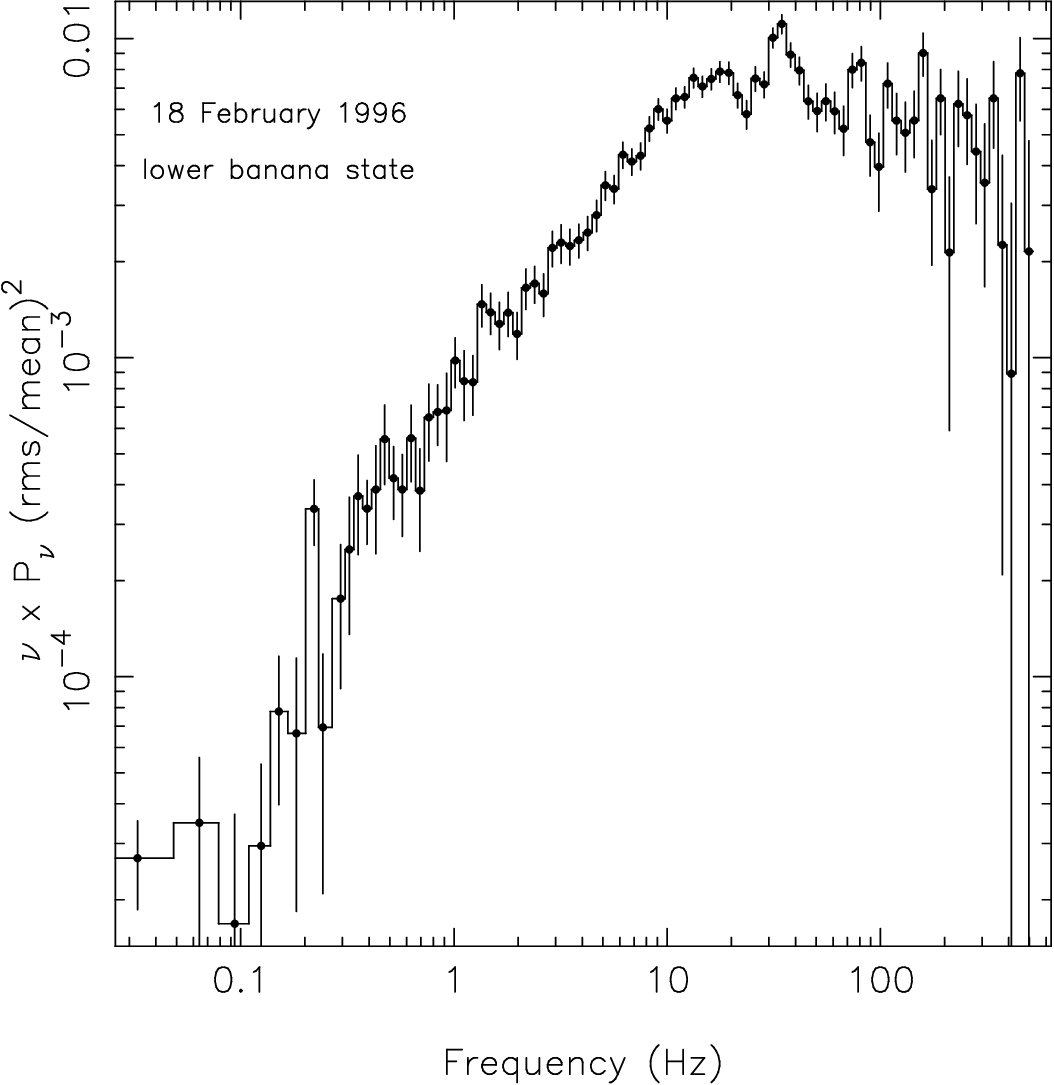}
\epsscale{0.45} \plotone{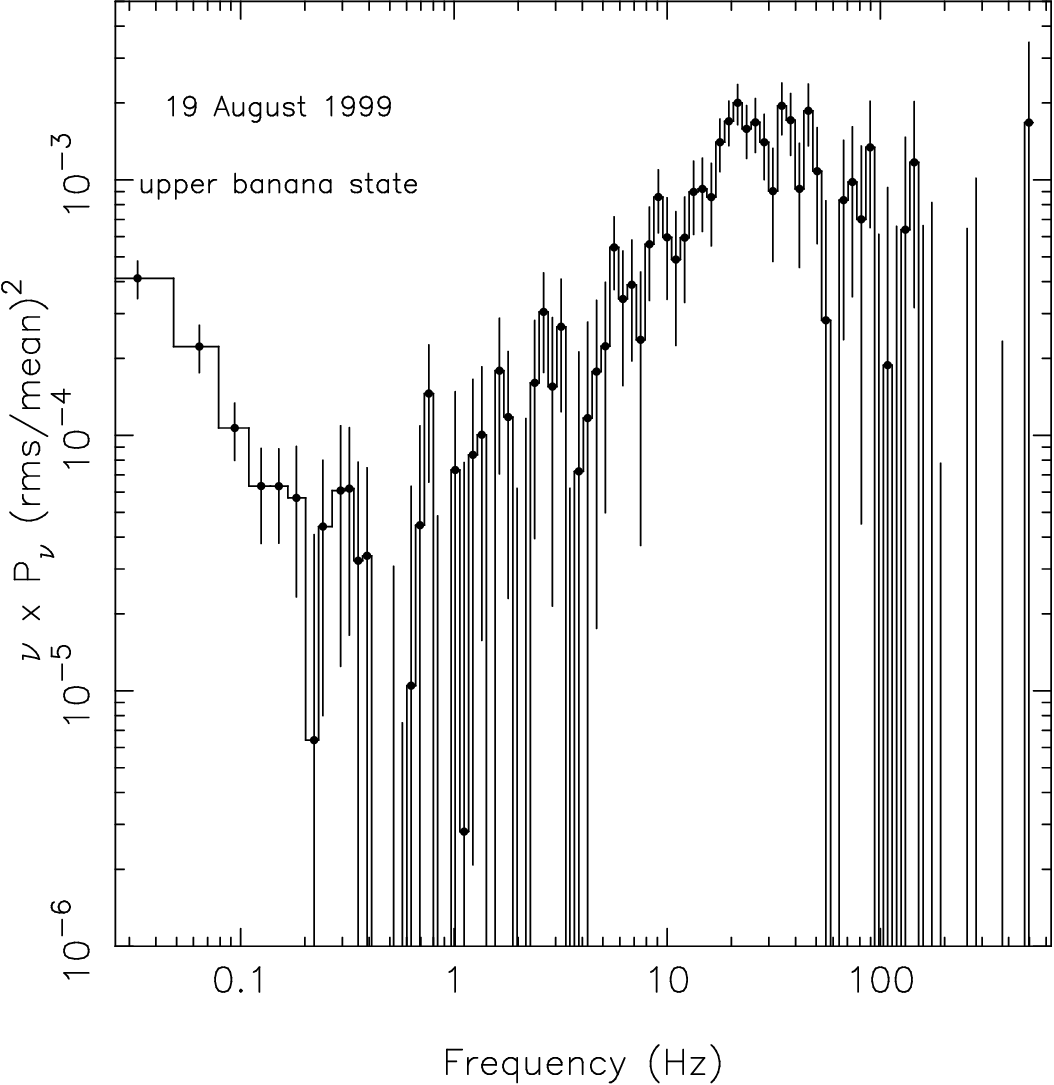} \caption{
Power spectral density for the representative states of \4u: island state, 
7 June 2001 (upper left), lower banana state,  18 February 1996 (upper
right) and upper banana,  19 August 1999 (lower). 
} 
\label{pds}
\end{figure}

\clearpage

\begin{figure}
\epsscale{0.85} 
\plotone{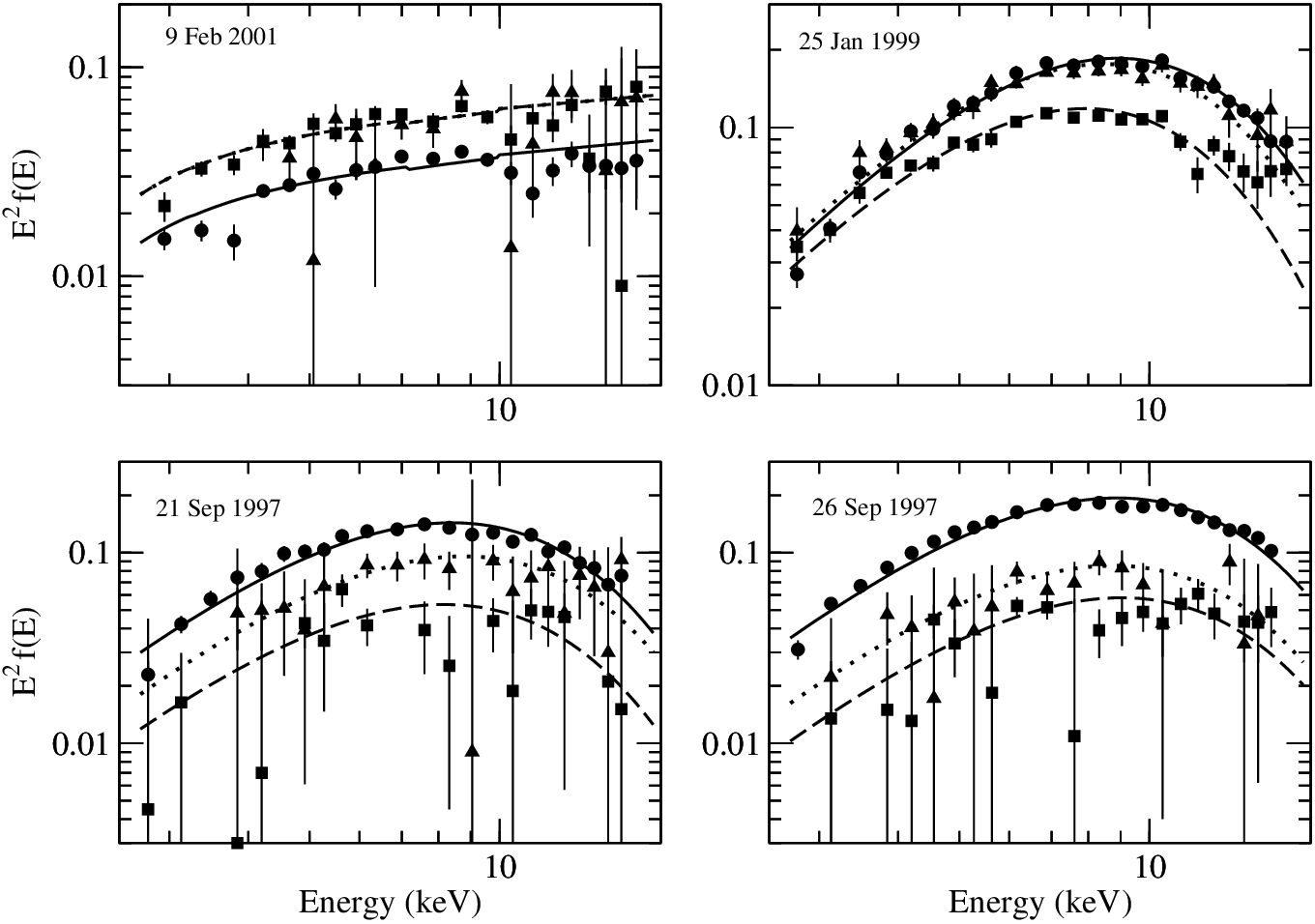}
\caption{Fourier resolved
spectra and their corresponding spectral-model fits for the days 
annotated in the figure.
The different frequency bands are as follows: circles $0.008 - 0.8$
Hz; squares $0.8 - 8$ Hz; triangles $8 - 64$ Hz. The solid,
dashed, and dotted curves are the respective model fits
plotted in $\nu f_{\nu}$ space. } \label{frs}
\end{figure}


\clearpage

\begin{figure}
\epsscale{0.85} 
\plotone{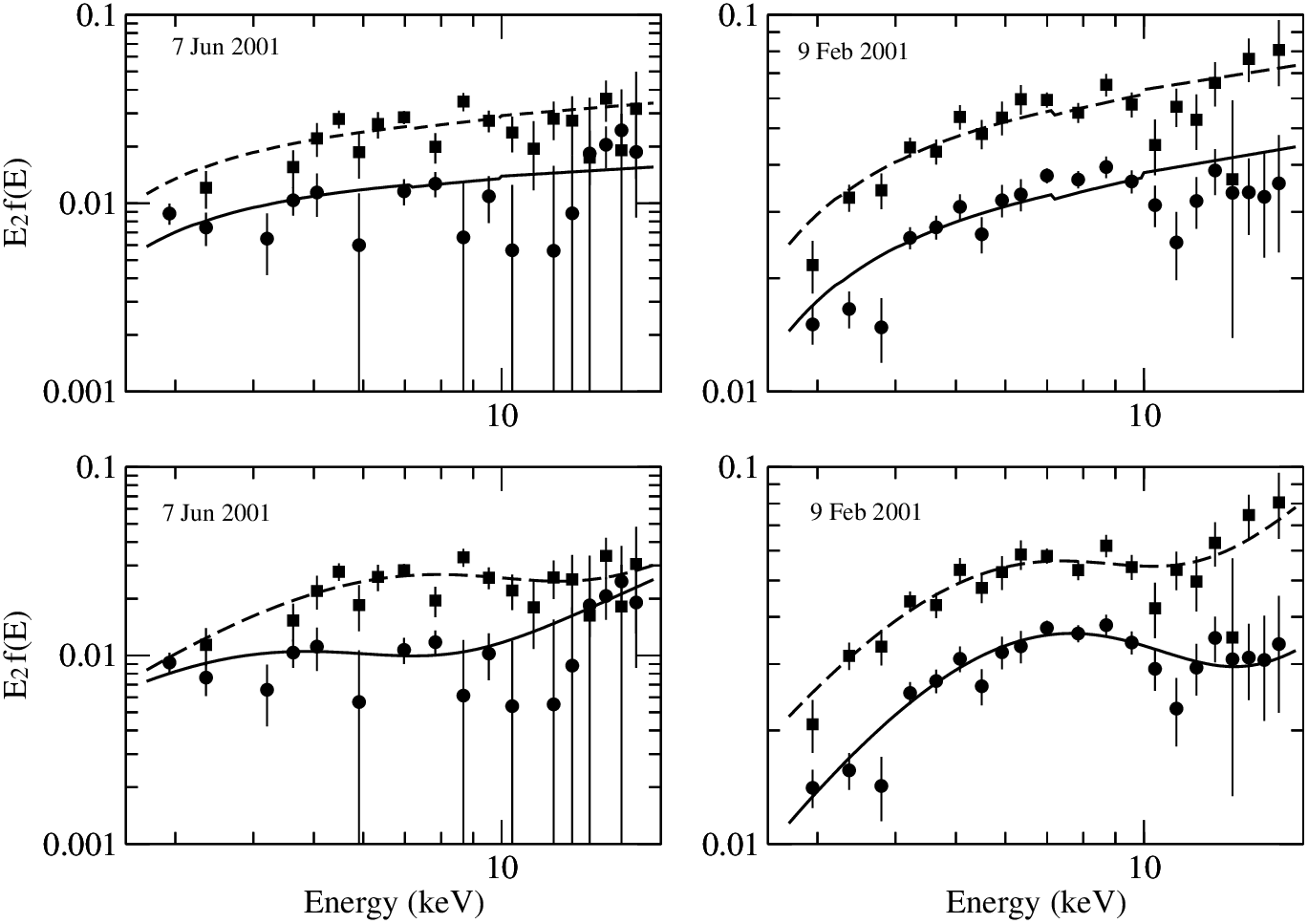} \caption{
 The upper figures show the low- and medium-frequency spectra for 
(a) the June 7, 2001 and
(b) February 9, 2001 island state observations fitted to an absorbed
powerlaw model (refer to \ref{ISplfits} for the parameter values).
The solid curves are the model in $\nu f_\nu$ space with the 
FRS spectral data points overlayed.
The lower figures show the spectral decomposition
of the same two data
sets with an additive blackbody plus powerlaw model.  } \label{island}
\end{figure}

\clearpage


\clearpage

\begin{thebibliography}{999}

\bibitem[di Salvo et al.(2001)]{salv01}
di Salvo, T., M\'endez, M., van der Klis, M., Ford, E., Robba, N.R. 2001,
ApJ, 546, 1107

\bibitem[Ford \& van der Klis(1998)]{ford98}
 Ford, E., \& van der Klis, M., 1998, ApJ, 506, L39

\bibitem[Foster etal.(1986)]{fost86}
Foster, A. J., Fabian, A. C., Ross, R. R. 1986, MNRAS, 221, 409

\bibitem[Gilfanov et al.(2000)]{gilf00}
Gilfanov, M., Churazov, E. \& Revnivtsev, M. 2000, MNRAS, 316, 923

\bibitem[Gilfanov et al.(2003)]{gilf03}
    Gilfanov, M., Revnivtsev, M., \& Molkov, S.,
  A\&A, 410, 217

\bibitem[Gilfanov \& Revnivtsev(2005)]{gilf05}
Gilfanov, M., \& Revnivtsev, M. 2005, Astr, Nachr., 326, 812

\bibitem[Kazanas \& Hua(1999)]{kazhua}
Kazanas, D. \& Hua, X.-M., 1999, ApJ, , 519, 750

\bibitem[Mendez \& van der Klis(1999)]{mend99}
 Mendez, M., \& van der Klis, M., 1999, ApJ, 517, L51.

\bibitem[Migliari et al.(2003)]{migl03}
Migliari et al., 2003, MNRAS, 342, L67

\bibitem[Narita et al (1999)]{nar99}
Narita, T., Grindlay, J.E., and Barret, D., bull AAS, 31, 904

\bibitem[Papadakis et al. (2005)]{papa05}
Papadakis, I. E., Kazanas, D., \& Akylas, A.  2005, ApJ, 631, 727

\bibitem[Papadakis (2006)]{papa06}
Papadakis, I.E., Ioannou, Z., \& Kazanas, D.,
 2006, ASNA, 327, 1047

\bibitem[Piraino et al (2000)]{pira00}
 Piraino, S., Santangelo, A., \& Kaaret, P., 2000, A\&A, 360, L35

\bibitem[Reig et al. (2006)]{reig06}
Reig, P., Papadakis, I.E., Shrader, C.R.,
 \&  Kazanas, D., 2006, ApJ, 644, 424

\bibitem[Revnivtsev et al. (1999)]{revn99}
Revnivtsev, M., Gilfanov, M., \& Churazov, E.  1999, A\&A, 347, L23

\bibitem[Revnivtsev et al. (2000)]{revn00} 
Revnivtsev, Mikhail G., 
Borozdin, Konstantin N., Priedhorsky, William C.,\& Vikhlinin, Alexey
2000, Ap.J., 230, 955

\bibitem[Revnivtsev et al.(2001)]{revn01}
Revnivtsev, M., Gilfanov, M., \& Churazov, E.  2001, A\&A, 380, 502

\bibitem[Revnivtsev \& Gilfanov (2006)]{revn06}
Revnivtsev, M. G., Gilfanov, M. R. 2006, A\&A, 453, 253

\bibitem[Sobolewska, M. A.; Zycki, P. T. (2006)]{Sob06}
Sobolewska, M.A., \& Zycki, P.T., 2006, MNRAS, 370, 405

\bibitem[Titarchuk \& Shaposhnikov (2005)]{tita05}
Titarchuk, L. \& Shaposhnikov, N. 2005, ApJ, 626, 298

\bibitem[van der Klis (2000)]{klis00}
 van der Klis, M., 2000, ARA\&A, 38, 717

\bibitem[van der Klis (2006)]{klis06}
 van der Klis, M., 2006,
 in "In: Compact stellar X-ray sources", Ed. by W. Lewin \&
  M. van der Klis. Cambridge Astrophysics Series, No. 39. Cambridge
  University Press.

\bibitem[Vignarca et al.(2003)]{vign03}
Vignarca, F., Migliari, S., Belloni, T., Psaltis, D., van der Klis, M.
2003, A\&A, 397, 729

\bibitem[Wijnands (2001)]{wijn01}
Wijnands, R., 2001, AdSpR, 28, 469

\bibitem[Zycki et al (2003)]{zycki03}
 Zycki, P., 2003, ASPC, 290, 135

\end{thebibliography}
\end{document}